\pdfoutput=1
\documentclass[journal]{IEEEtran}
\usepackage{cite}
\usepackage{amsmath,amssymb,amsfonts}
\usepackage{graphicx}
\usepackage{textcomp}
\usepackage[normalem]{ulem}
\usepackage{booktabs}
\usepackage{makecell}
\usepackage{multirow}
\usepackage{todonotes}

\usepackage{threeparttable}
\usepackage{subcaption}
\usepackage{tablefootnote}
\usepackage{hyperref}
\usepackage{url}
\usepackage[ruled, lined, linesnumbered, commentsnumbered, longend]{algorithm2e}
\DontPrintSemicolon
\usepackage{algpseudocode}
\usepackage[switch, modulo]{lineno}
\usepackage{color}

\def\BibTeX{{\rm B\kern-.05em{\sc i\kern-.025em b}\kern-.08em
    T\kern-.1667em\lower.7ex\hbox{E}\kern-.125emX}}
\begin{document}

\title{Federated Multi-organ Segmentation with Inconsistent Labels}

\author{Xuanang Xu, Hannah H. Deng, Jaime Gateno, and Pingkun Yan,~\IEEEmembership{Senior Member, IEEE}
\thanks{Asterisk indicates corresponding author.}%
\thanks{X.~Xu and P.~Yan* are with the Department of Biomedical Engineering and the Center for Biotechnology and Interdisciplinary Studies at Rensselaer Polytechnic Institute, Troy, NY 12180 USA (xux12@rpi.edu, yanp2@rpi.edu).}%
\thanks{H.H.~Deng and J.~Gateno are with the Department of Oral and Maxillofacial Surgery at Houston Methodist Research Institute, Houston, TX 77030 USA (hdeng@houstonmethodist.org, jgateno@houstonmethodist.org).}
\thanks{J.~Gateno is also with the Department of Surgery (Oral and Maxillofacial Surgery) at Weill Medical College at Cornell University, New York, NY 10065 USA.}
\thanks{This work was supported in part by the National Science Foundation (NSF) under CAREER award OAC 2046708 and the National Institutes of Health (NIH) under awards R01DE022676 and R21EB028001.}
}

\maketitle

\begin{abstract}
Federated learning is an emerging paradigm allowing large-scale decentralized learning without sharing data across different data owners, which helps address the concern of data privacy in medical image analysis. However, the requirement for label consistency across clients by the existing methods largely narrows its application scope. In practice, each clinical site may only annotate certain organs of interest with partial or no overlap with other sites. Incorporating such partially labeled data into a unified federation is an unexplored problem with clinical significance and urgency. This work tackles the challenge by using a novel federated multi-encoding U-Net (Fed-MENU) method for multi-organ segmentation. In our method, a multi-encoding U-Net (MENU-Net) is proposed to extract organ-specific features through different encoding sub-networks. Each sub-network can be seen as an expert of a specific organ and trained for that client. Moreover, to encourage the organ-specific features extracted by different sub-networks to be informative and distinctive, we regularize the training of the MENU-Net by designing an auxiliary generic decoder (AGD). Extensive experiments on six public abdominal CT datasets show that our Fed-MENU method can effectively obtain a federated learning model using the partially labeled datasets with superior performance to other models trained by either localized or centralized learning methods. Source code is publicly available at \url{https://github.com/DIAL-RPI/Fed-MENU}.
\end{abstract}

\begin{IEEEkeywords}
Federated learning, Deep learning, Medical image segmentation, Inconsistent labels.
\end{IEEEkeywords}

\section{Introduction}
\label{sec:introduction}

\IEEEPARstart{C}{onvolutional} neural network (CNN) based deep learning (DL) as a data-driven methodology has demonstrated unparalleled performance in various segmentation tasks, providing that the model can train on large-scale data with sufficient diversity. To suffice the large appetite of CNNs, researchers often collect data from multiple sources to jointly train a model for improved performance. However, in the healthcare domain, such \emph{centralized learning} paradigm is often impractical because the clinical data cannot be easily shared across different institutions due to the regulations, such as Health Insurance Portability and Accountability Act (HIPAA). To overcome the barrier of data privacy and realize large-scale DL on isolated data, federated learning (FL)~\cite{Mcmahan2017FedAvg}, an emerging \emph{decentralized learning} paradigm, has been adopted in the medical image analysis, solving various clinical problems such as prostate segmentation~\cite{Roth2021FedProstateSupernet,Sarma2021FedProstate} and COVID-19 diagnosis~\cite{Yang2021FedSemiCovid,Dayan2021FedCovid}. 

FL allows different data-owners to collaboratively train one global DL model without sharing the data. The model training is completed by iterating over a server node and several client nodes. Each client individually trains a copy of the global model using their local data after the server updates the global model by aggregating the locally trained models. By repeating this process, the global model can effectively absorb the knowledge contained in the client datasets without data sharing. As data privacy has become a critical issue concerned by healthcare stakeholders, FL attracts a growing attention from both the research and clinical communities in recent years. 

Although the capability of FL for medical image analysis has been demonstrated by the prior studies, it comes with limitations. A critical issue is about the strict requirement for the label consistency. Specifically, in an FL system, all the participating sites need to have identical regions-of-interest (ROIs) annotated on their local data, so the FL model can be optimized following the same objective across different clients. However, in practical scenarios, different clinical sites often have different expertise and thus follow different protocols for data annotation. This leads to inconsistent ROI labels across different sites. The requirement for labeling consistency hinders the FL methods from large-scale deployment in practice. Therefore, a more flexible FL framework supporting the training with inconsistently labeled client data is highly desired. From a technical point of view, this is an FL problem with partial labels since each client merely has a partially labeled dataset with respect to the whole set of ROIs in the federation. To the best of our knowledge, this is a new problem that has not yet been explored before, but at the same time is of great clinical significance and technical urgency to tackle.

In this paper, we propose a novel method, called federated multi-encoding U-Net (Fed-MENU), to address the above challenge. We then demonstrate its performance on a representative task, \emph{i.e.}, multiple abdominal organ segmentation from computed tomography (CT) images. The underlying assumption of our design is that, although the data from different clients show disparities in terms of the labeled ROIs, they share the same or similar image contents and thus can provide complementary information to facilitate the learning of robust features in a unified FL system. Since the client datasets are partially labeled with different ROIs, they can be seen as a set of experts with different expertise. Each expert focuses on learning the features within its expertise (labeled ROIs) while avoiding introducing the unreliable or noisy information from the non-expertise (unlabeled ROIs). To achieve this goal, we design a multi-encoding U-Net (MENU-Net) to decompose the multi-organ feature learning task into a series of individual sub-tasks of organ-specific feature learning. Each of them is bound with a sub-network in the MENU-Net. During the local training stage of FL, the client node can only tune the sub-network associated with the organ on which they have labels (expertise) while keeping the other parts of the network unchanged. Due to this decomposed feature learning strategy, the MENU-Net is encouraged to learn organ-specific features via different sub-networks without interference from other label-absent organs. Moreover, to further encourage the extracted organ-specific features to be informative and distinctive from other organs, we design an auxiliary generic decoder (AGD) to regularize the training of the MENU-Net. As a result, the MENU-Net can extract discriminative features during its encoding stage and thus facilitate the segmentation in the following decoding stage.

To demonstrate the performance of our Fed-MENU method, we conducted extensive experiments using four public abdominal CT datasets, each of which is annotated with a different set of abdominal organs. Our results show that, without sharing the raw data, the proposed Fed-MENU method can effectively utilize the isolated datasets with different partial labels to train a global model for multi-organ segmentation. The performance of the trained model is superior to the localized learning model trained by any single dataset and also the centralized learning model trained by combining all the datasets.
The main contributions of this paper are three-fold.
\begin{itemize}
    \item
    We addressed a new problem in FL to enable collaboratively training of a global model using isolated datasets with inconsistent labels.
    \item
    We proposed a novel Fed-MENU method to deal with this challenging problem. In our method, organ-specific features are individually extracted by different sub-networks of the MENU-Net and further enhanced by the accompanied AGD.
    \item
    We evaluated the performance of our Fed-MENU method for multi-organ segmentation using six public abdominal CT datasets. Our experimental results on both in-federation and out-of-federation testing sets demonstrated the effectiveness and superior performance of our design.
\end{itemize}

The rest of this paper is organized as follows. Section~\ref{sec:related_works} gives a brief review of the previous literature related to this work. Section~\ref{sec:method} illustrates the details of the proposed method. Section~\ref{sec:experiments} presents the experiments and results on four public datasets. Finally, we discuss the limitation of this work and conclude it in Section~\ref{sec:conclusion}.

\section{Related works}
\label{sec:related_works}

Since our study involves both FL-based medical image analysis and DL-based image segmentation with partially labeled data, we review the related works in these two areas before presenting our proposed method.

\subsection{Background of federated learning}
FL, first emerged in 2017~\cite{Mcmahan2017FedAvg}, is a decentralized learning paradigm designed to address data privacy issue in deep learning. Unlike the conventional centralized deep learning that requires all training data gathered on a server node, FL allows distributed data owners to collaboratively train a model without sharing their data. Federated averaging (FedAvg)~\cite{Mcmahan2017FedAvg} was acknowledged as the benchmarking algorithm in FL, which defined the framework for the following studies in this field. In FedAvg, the model is simultaneously trained on several client devices and the clients transmit the trained model (parameters) to the server to update/maintain the global model (parameters). A major challenge in FL is how to deal with the client data that is not independently and identically distributed (non-iid) or heterogeneous~\cite{Hsu2019FedAvgM,Li2020FedProx,Acar2021FedDyn,Li2021FedBN,Bernecker2022FedNorm,Xu2022FedSM,Xu2022FedCross}. For example, Li \emph{et al.}~\cite{Li2020FedProx} derived FedProx from FedAvg by regularizing the local models to be closer to the global model, in which way the method showed more stable and accurate performance when learning over the heterogeneous data. Acar \emph{et al.}~\cite{Acar2021FedDyn} proposed FedDyn by introducing a dynamic regularization term to the local training objective of FedAvg. The resulting model achieved not only better robustness against data heterogeneity but also higher communication efficiency. Li \emph{et al.}~\cite{Li2021FedBN} proposed a simple yet effective scheme, FedBN, to address the heterogeneity issue in feature space by keeping the batch normalization layers locally updated. Except for the above FL algorithms, data harmonization~\cite{Nan2022Harmonisation} can also be considered as a way to address the heterogeneities of multi-center data in FL.

\subsection{Federated learning for medical image analysis}

In recent years, FL has been widely applied in various medical image analysis tasks to address the conflict between large-scale DL model training and healthcare data privacy. For example, Dayan \emph{et al.}~\cite{Dayan2021FedCovid} trained an FL model for COVID-19 clinical outcome prediction using chest X-ray images collected from 20 institutes across the globe. The experimental results showed that, by utilizing multi-site data, the trained FL model achieved an average area under the curve (AUC) around 0.920 for predicting the patients’ oxygen therapy categories after 24- and 72-hour periods from initial admission to the emergency department. This AUC value of the FL model is 16\% higher than the average AUC (0.795) of the 20 locally trained models with single-site data. Xia \emph{et al.}~\cite{Xia2021AutoFedAvg} derived an auto-FedAvg method from FedAvg algorithm~\cite{Mcmahan2017FedAvg} for medical image segmentation using data that are not independently and identically distributed (non-iid). Their method automatically learns the aggregation weights of each client based on the client data distribution. Roth \emph{et al.}~\cite{Roth2021FedProstateSupernet} combined FL with neural architecture search strategy to develop a super-network with a self-learned structure for whole prostate segmentation from multi-institute magnetic resonance imaging (MRI) data. Recently, Yang \emph{et al.}~\cite{Yang2021FedSemiCovid} extended FL to semi-supervised learning paradigm and applied it to COVID region segmentation using chest CT data from three nations. Another work by Park \emph{et al.}~\cite{Park2021FLTransformer} employed the Vision Transformer (ViT) architecture~\cite{Dosovitskiy2020ViT} in an FL framework to diagnose the COVID-positive cases from chest X-ray images.

The prior studies~\cite{Roth2020FedBreast,Liu2021FedDG,Dayan2021FedCovid,Xia2021AutoFedAvg,Roth2021FedProstateSupernet,Yang2021FedSemiCovid,Jiang2022IOP-FL,Guo2022Auto-FedRL,Xu2022FedCross} have demonstrated the feasibility and effectiveness of FL in solving the problem of large-scale DL model training without data sharing in the healthcare community. However, these successes are built upon a prerequisite that the participating sites have consistently labeled data so that they can contribute to the same training objective of one global model. However, such a strong requirement may not be met in the real scenarios, where different clinical sites often follow different protocols to annotate their local data. Although some prior efforts like FedMix proposed by Wicaksana \emph{et al.}~\cite{Wicaksana2022FedMix} allowed mixed supervised FL with different levels of label (\emph{e.g.}, pixel-level mask, bounding-box-level annotation, and image-level labels), the gap caused by the label inconsistency (in terms of the target classes) remained unresolved, which largely narrows the application scope of the FL-based methods for medical image analysis and thus motivates our study in this work.

It is worth noting that Dong \emph{et al.}~\cite{Dong2022PartialLabelFL-MICCAI,Dong2022PartialLabelFL-TMI} worked on a similar problem of federated learning-based X-ray image classification with partially labeled data, which is related to our work and thus provided positive implications for us during the revision of this paper. However, there are significant differences between these two works. First of all, Dong \emph{et al.}’s work is targeting at the image classification task while our work focused on the image segmentation task, which is more general and common in terms of the scope of clinical application. As a consequence, the methods proposed in these two works are fundamentally different and thus cannot be directly compared. Moreover, the method by Dong \emph{et al.} was only evaluated in an in-federation setting, \emph{i.e.}, the testing data follows the distribution of the training data from one of the clients, while our study also includes an out-of-federation evaluation, \emph{i.e.}, the testing data comes from an unseen domain out of any one of the client datasets, which is more exhaustive and challenging.

\subsection{Medical image segmentation with partial labels}

Due to the high cost of data annotation, medical images are often partially labeled with different ROIs or labels, even though they may share the same imaging field. This created a barrier for the DL-based medical image segmentation methods~\cite{Shi2022NCseg,Ye2022NCseg} when the model is trained on the partially labeled data. To solve this problem, Yan \emph{et al.}~\cite{Yan2020PartialLabel} developed a universal lesion detection algorithm trained by CT images labeled with different lesion types. In their method, the missing labels were mined by exploiting clinical prior knowledge and cross-dataset knowledge transfer. Recently, Petit \emph{et al.}~\cite{Petit2021PartialLabel} proposed to conduct partially labeled DL training through an iterative confidence relabeling method, in which a self-supervised scheme was employed to iteratively relabel the missing organs by introducing pseudo labels into the training set. For the partial label problem in segmentation tasks, Fang and Yan~\cite{Fang2020PIPOFAN} proposed a pyramid-input-pyramid-output feature abstraction network (PIPO-FAN) for multi-organ segmentation, in which a target adaptive loss is integrated with a unified training strategy to enable image segmentation over multiple partially labeled datasets with a single model. Shi \emph{et al.}~\cite{Shi2021MarginalLoss} designed two types of loss functions, namely marginal loss and exclusion loss, to train a multi-organ segmentation network using a set of partially labeled datasets. Furthermore, Zhang \emph{et al.}~\cite{Zhang2021PartialLabel} proposed an ingenious approach, namely dynamic on-demand network (DoDNet), to achieve multi-organ segmentation with partially labeled data. Unlike the conventional deep neural networks with fixed parameters after training, the DoDNet can dynamically generate model parameters to adapt to different organ segmentation tasks. This innovative design largely improved the model efficiency and flexibility.

Although the problem of training DL models using partially labeled data~\cite{Yan2020PartialLabel,Petit2021PartialLabel,Fang2020PIPOFAN,Shi2021MarginalLoss,Zhang2021PartialLabel,Liu2021PartialLabel} had been studied in the context of centralized learning, it is still an unexplored area for the field of FL. In this paper, we present the Fed-MENU method, which can learn from partially labeled data distributed on different sites.

\section{Method}
\label{sec:method}

\begin{figure*}
	\centering
	\includegraphics[width=\textwidth]{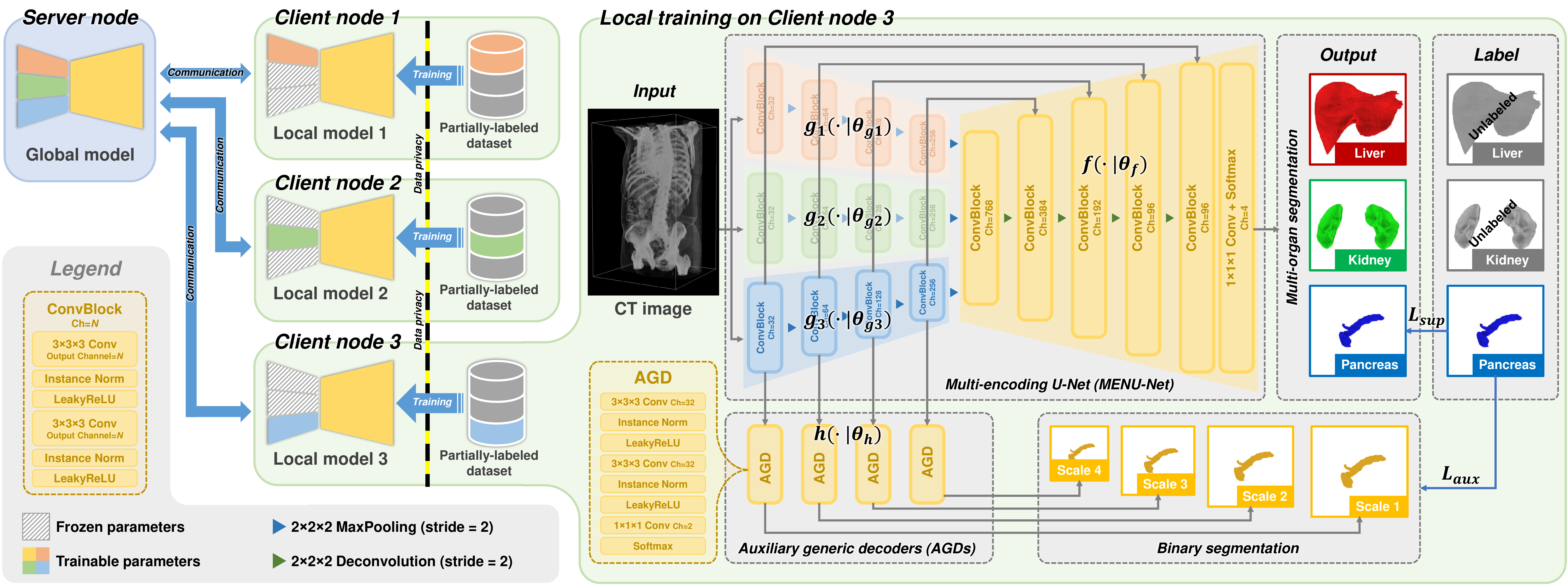}
	\caption{Scheme of the proposed Fed-MENU method. For better understanding, the method is presented using an example scenario where three clinical sites with partially labeled abdominal CT images collaboratively train an FL model for multi-organ segmentation of the liver, kidney, and pancreas. The left panel of the figure shows the federation structure. The right panel details the local training procedure on Client node 3, where the pancreas is the only labeled organ in this client dataset. For brevity, we present only the connections between the third sub-encoder (colored in blue) and the shared decoder, as well as the AGDs. In the final implementation, all sub-encoders are connected to the shared decoder and AGDs at equivalent positions.}
	\label{fig:1}
\end{figure*}

Fig.~\ref{fig:1} gives an overview of the proposed Fed-MENU method. For a better understanding, it is presented in a specific scenario where the federation contains three client nodes and each node possesses a CT dataset partially labeled with one of the abdominal organs (liver, kidney, and pancreas). We then describe the FL framework adopted by our method in Section~\ref{sec:method:FL}. The technical innovation of this work resides in the MENU-Net designed for organ-specific feature extraction (Section~\ref{sec:method:net}) and the AGD designed for organ-specific feature enhancement (Section~\ref{sec:method:agd}). The training configurations and other technical details of the proposed method are provided in Section~\ref{sec:method:details}.

\subsection{Federated learning framework}
\label{sec:method:FL}

Our method inherits the framework of the FedAvg~\cite{Mcmahan2017FedAvg} algorithm, which is a commonly adopted benchmark in FL. It consists of one server node and $K$ client nodes. The server takes the responsibility of coordinating the communication and computation across different clients, while the clients focus on model training using their local data and devices.
Given the universal label set of $M$ organs, each of the client nodes has a local dataset $D_k$ partially labeled with a subset of $N_k$($\le$$M$) organs.
Our goal is to train a $M$-organ segmentation network $F(\cdot,\theta)$ using the partially labeled datasets $\{D_k\}_{k=1}^K$, which reside in different client nodes and cannot be shared or gathered for centralized training. 

The training process of the FL framework consists of $T$ rounds of communication between the server and the client nodes. Specifically, in the $t$-th communication round, each client node $k$ will first download the parameters of the current segmentation network $F(\cdot,\theta^t)$ on server (denoted as global model), resulting in a local copy of $F(\cdot,\theta^t)$ (denoted as local model). The client node then trains the local model using its local dataset $D_k$ and device for $E$ epochs. After the local training, the server collects the trained local models $F(\cdot,\theta_k^t)$ from all the $K$ client nodes and aggregates them into a new global model through a parameter-wise averaging: 
\begin{equation}
    \theta^{t+1} = \sum_k^K \frac{|D_k|}{\sum_j^K|D_j|}\cdot\theta_k^t,
    \label{eq:1}
\end{equation}
where $|D_k|$ indicates the sample size of dataset $D_k$. The whole FL procedure is accomplished by repeating the above process until the global model $F(\cdot,\theta^t)$ converges.

Because the local models are individually trained on the client nodes, the server node only needs to transfer the model parameters instead of the raw data from the clients. That helps the FL model obtain knowledge contained in the isolated client datasets without violating the data privacy. However, since the original FedAvg framework is designed for FL with fully labeled data, there is a technical gap to bridge before we can deploy it to the case of partially labeled data. Therefore, we propose the following MENU-Net with AGD to meet the need.

\subsection{MENU-Net for organ-specific feature extraction}
\label{sec:method:net}

A straightforward way to train a multi-organ segmentation network using the partial labels is to calculate the training loss merely using the labeled organs while ignoring the unlabeled ones. However, in the case of FL with partial labels, different organs are labeled in the client datasets, focusing on different expertise for image segmentation. As a consequence, the segmentation performance of the trained local models would be biased to the labeled organs. Intuitively, a uniform aggregation of all local models (see Eq.~\ref{eq:1}) may weaken the expert client models. A more reasonable way may be to promote the strength of the local models on segmenting the labeled organs and avoid the interference from their weaknesses with the unlabeled organs.

To achieve the above goal, we designed the MENU-Net to explicitly decompose multi-organ feature extraction into several individual processes of organ-specific feature extraction. As illustrated in the right panel of Fig.~\ref{fig:1}, the MENU-Net consists of $M$ sub-encoders $\{g_m(\cdot\,|\,\theta_{g_m})\}_{m=1}^{M}$ followed by a shared decoder $f(\cdot\,|\,\theta_{f})$ as well as the skip connections between them. Fig.~\ref{fig:1} shows an example with $M$=3. Each sub-encoder $g_m(\cdot\,|\,\theta_{g_m})$ serves as an organ-specific feature extractor for the $m$-th organ. An input image $x$ is fed to all the sub-encoders to get the features of all the organs in parallel. The extracted features are then concatenated into one stream and fed to the shared decoder $f(\cdot\,|\,\theta_{f})$, which is used to interpret the organ-specific features into the multi-organ segmentation masks. This process is formally defined as
\begin{align}
    F(x) &= f( (g_1(x) \oplus g_2(x) \ldots \oplus g_M(x)) ) \nonumber \\ 
    &= f \left( \oplus_{m=1}^{M} g_m(x) \right),
\end{align}
where $\oplus$ denotes channel-wise concatenation. 

Given a client node with partial labels on the $m$-th organ, the local model training only tunes the parameters in the sub-encoder $g_m(\cdot\,|\,\theta_{g_m})$ and the shared decoder $f(\cdot\,|\,\theta_{f})$, which can be expressed as:
\begin{equation}
    \underset{\theta_f,\theta_{g_m}\subset\theta}{\arg\min}\ L_{sup}\left[ F_m(x\,|\,\theta),\ \hat{y}_m \right],
\end{equation}
where $F_m(x)$ and $\hat{y}_m$ are the predicted and ground-truth segmentation of the $m$-th organ, respectively. $L_{sup}$ denotes the supervised loss measuring the similarity between the prediction and ground truth, which can be in any form of pixel-wise loss functions (\emph{e.g.}, Cross-entropy loss and Dice loss~\cite{Milletari2016VNet}) or their combinations.

\begin{table*}[t]
	\caption{Detailed information of client datasets involved in this study. (L: liver, K: kidney, P: pancreas, S: spleen, G: gallbladder, \checkmark: labeled organ, $\times$: unlabeled organ, OoF: out-of-federation)}
	\centering
	\begin{threeparttable}[b]
		\begin{tabular}{p{20pt}<{\centering}|p{54pt}<{}p{32pt}<{\centering}p{38pt}<{\centering}p{32pt}<{\centering}p{32pt}<{\centering}|p{4pt}<{\centering}p{4pt}<{\centering}p{4pt}<{\centering}p{4pt}<{\centering}p{4pt}<{\centering}|p{90pt}<{}}
			\toprule[2pt]
			\multirow{4}*{\makecell[c]{Datasets}}
			~&\multicolumn{5}{c|}{Image information}&\multicolumn{5}{c|}{Labeled organ}&\multirow{4}*{\makecell[l]{Imaging phase}}\\
			\cmidrule[0.5pt]{2-11}
			~&\# of CT images (train/val./test)&Slice size [in pixel]&\# of slice [per image]&Spacing [$mm$]&Thickness [$mm$]&L&K&P&S&G\\
			\midrule[1pt]
			\multicolumn{1}{l|}{Client \#1: Liver}&131 (79/13/39)&512&74$\sim$987&0.56$\sim$1.00&0.70$\sim$5.00&$\checkmark$&$\times$&$\times$&$\times$&$\times$&Portal-venous phase\\
			\multicolumn{1}{l|}{Client \#2: Kidney}&210 (126/21/63)&\{512,796\}&29$\sim$1059&0.44$\sim$1.04&0.50$\sim$5.00&$\times$&\checkmark&$\times$&$\times$&$\times$&Late arterial contrast phase\\
			\multicolumn{1}{l|}{Client \#3: Pancreas}&281 (169/28/84)&512&37$\sim$751&0.61$\sim$0.98&0.70$\sim$7.50&$\times$&$\times$&\checkmark&$\times$&$\times$&Portal-venous phase\\
			\multicolumn{1}{l|}{Client \#4: Spleen}&41 (24/5/12)&512&31$\sim$168&0.61$\sim$0.98&1.50$\sim$8.00&$\times$&$\times$&$\times$&\checkmark&$\times$&Portal-venous phase\\
			\multicolumn{1}{l|}{Client \#5: AMOS}&200 (120/20/60)&60$\sim$768&64$\sim$512&0.45$\sim$3.00&0.82$\sim$5.00&\checkmark&\checkmark&\checkmark&\checkmark&\checkmark&See Appendix B.1 in \cite{Ji2022AMOS}\\
			\multicolumn{1}{l|}{OoF client: BTCV}&30 (0/0/30)&512&85$\sim$198&0.59$\sim$0.98&2.50$\sim$5.00&\checkmark&\checkmark&\checkmark&\checkmark&\checkmark&Portal-venous phase\\
			\midrule[0.5pt]
			\multicolumn{1}{l|}{Total}&893(518/87/288)&60$\sim$796&29$\sim$1059&0.44$\sim$3.00&0.50$\sim$8.00&-&-&-&-&-&-\\
		\bottomrule[2pt]
		\end{tabular}
	\end{threeparttable}
	\label{tab:1}
\end{table*}

\subsection{AGD for organ-specific feature enhancement}
\label{sec:method:agd}

Benefitting from the decomposed feature extraction in MENU-Net, organ-specific knowledge from the labeled clients (experts) can be individually learned by different sub-encoders with less interference from other unlabeled clients (non-experts).
However, since each local model can only see the images from one client dataset, the organ-specific features learned by the sub-encoders may also include some domain-specific information.
This is unfavorable for the subsequent shared decoder.
Ideally, the sub-encoder should focus on the structural information of the corresponding organ, which is invariant across domains. In addition, the extracted organ-specific features should be distinctive enough from that of other organs, making the subsequent shared decoder easier to interpret them into different organs' segmentation.

Motivated by the above idea, we design the AGD shared across different sub-encoders to help regularize the training of our MENU-Net, aiming to enhance the extracted organ-specific features.
Specifically, as illustrated in the mid-bottom of Fig.~\ref{fig:1}, given the organ-specific features $g_m(x)$ extracted by an arbitrary sub-encoder $g_m(\cdot\,|\,\theta_{g_m})$, we feed them to a set of AGDs to perform an organ-agnostic (binary) segmentation of the $m$-th organ:
\begin{equation}
    G_m(x)=h(g_m(x)).
\end{equation}
Here, we denote the collection of all AGDs as $h(\cdot\,|\,\theta_{h})$ parameterized by $\theta_{h}$. 
The AGD has a lite structure consisting of three convolutional layers. The first two convolutional layers contain 3$\times$3$\times$3 kernels followed by instance normalization~\cite{Ulyanov2016InstanceNorm} and LeakyReLU activation~\cite{Maas2013LeakyReLU}. The last convolutional layer has 1$\times$1$\times$1 kernels with two output channels followed by a softmax activation.
Due to the shallow structure of AGD, it has limited representation ability and thus enforces the preceding sub-encoder layers to extract more discriminative features to approach the organ segmentation.
On the other hand, since the AGD is working as a generic decoder for different organs when combined with different sub-encoders, the sub-encoders are encouraged to extract features distinctive enough from each other so that the following AGD can interpret them into the corresponding organ's segmentation without extra information.

During the local training stage, the AGDs are tuned together with different sub-encoders over all client nodes by minimizing the segmentation error between the output $G_m(x)$ and the corresponding ground truth $\hat{y}_m$. This process can be expressed as:
\begin{equation}
    \underset{\theta_h,\theta_{g_m}}{\arg\min}\ L_{aux}\left[ G_m(x\,|\,\theta_h,\theta_{g_m}),\ \hat{y}_m \right],
    \label{eq:5}
\end{equation}
where $L_{aux}$ denotes the auxiliary loss quantifying the binary segmentation error between the AGDs' output and the corresponding ground truth. We calculate $L_{aux}$ as a sum of the Cross-entropy loss and Dice loss~\cite{Milletari2016VNet}. Note that, since the AGDs are connected to the multi-scale levels of the sub-encoder, they have multiple outputs in different scales. All these outputs will be resampled to the original size of the ground-truth segmentation $\hat{y}_m$ and counted in the auxiliary loss in Eq.~\ref{eq:5}.
After the local training, the tuned parameters of AGDs from all clients are aggregated through the FedAvg algorithm as shown in Eq.~\ref{eq:1}, which is the same as that of the shared decoder $f(\cdot\,|\,\theta_{f})$.
The AGDs are used to regularize the training of the proposed MENU-Net. During the inference stage, we only utilize the trained MENU-Net to infer the multi-organ segmentation masks.

\subsection{Implementation details}
\label{sec:method:details}

The proposed method is implemented for 3D segmentation using PyTorch. Model parameters in the segmentation network are initialized using the Xavier algorithm~\cite{Glorot2010Xavier} and optimized by an SGD optimizer. The learning rate is initialized to be 0.01 and decayed throughout the training following a poly learning rate policy~\cite{Isensee2021nnUNet} with a momentum factor of 0.99. We train the model for $T$=400 rounds of communication ($E$=1 epoch of local training per round) and evaluate its performance on the validation set every epoch using Dice similarity coefficient (DSC) as the metric. The model achieving the highest DSC on the validation set is selected as the final model to be evaluated on the testing set. The training batch size is set to 4 on four NVIDIA A100 GPUs. The CT images are resampled to a uniform resolution of 1.0$\times$1.0$\times$1.5$mm^3$ (pixel spacing of 1.0$\times$1.0$mm^2$ and slice thickness of 1.5$mm$) before training. Image patches with a fixed pixel size of 256$\times$256$\times$32 are randomly cropped from the resampled CT images as the input of the segmentation network. The image intensities are normalized from [-200.0, 400.0] Hounsfield units (HU) to [0.0, 1.0] for a good soft-tissue contrast. Random translation ([-20,20]$mm$) and rotation ([-0.1,0.1]$rad$) are used to augment the training samples. In the inference time, an unseen CT image is first resampled to the resolution of 1.0$\times$1.0$\times$1.5$mm^3$ and divided into a series of patches, which are then fed to the trained segmentation network. The predicted patch-wise segmentation will be assembled to get the final segmentation mask, which is then resampled back to their original CT resolution for quantitative evaluation. Unless otherwise noted, all the competing methods and ablation models in our experiments are trained and evaluated using the same configuration as our method. For better reproducibility, the source code is released at \url{https://github.com/DIAL-RPI/Fed-MENU}. 

\section{Experiments}
\label{sec:experiments}

\subsection{Datasets}
We conducted experiments using six public abdominal CT image datasets, including 1) liver tumor segmentation challenge (LiTS) dataset~\cite{Bilic2019LiTS}\footnote{https://competitions.codalab.org/competitions/17094}, 2) kidney tumor segmentation challenge (KiTS) dataset~\cite{Heller2019kits19,Heller2020KiTS}\footnote{https://kits19.grand-challenge.org/home}, 3) pancreas and 4) spleen segmentation datasets in medical segmentation decathlon challenge~\cite{Simpson2019MSD,Antonelli2021MSD}\footnote{http://medicaldecathlon.com} (Task \#7 and \#9, respectively), 5) multi-modality abdominal multi-organ segmentation challenge (AMOS) dataset~\cite{Ji2022AMOS}\footnote{https://amos22.grand-challenge.org}, and 6) multi-atlas labeling beyond the cranial vault challenge (BTCV) dataset~\cite{Landman2015BTCV}\footnote{https://www.synapse.org/\#!Synapse:syn3193805/wiki/89480}. For brevity, we denote the six datasets as the liver, kidney, pancreas, spleen, AMOS, and BTCV, respectively, in the rest of the paper. Table~\ref{tab:1} shows the detailed information of these datasets.

\textbf{Liver dataset} contains 131 images, whose sizes range between [74$\sim$987]$\times$[512]$\times$[512] (in $D$$\times$$H$$\times$$W$ pixels). The in-plane spacing of these CT slices varies from 0.56$mm$ to 1.00$mm$ and the slice thickness varies from 0.70$mm$ to 5.00$mm$. Each CT image has a pixel-wise annotation of liver and tumor segmentation stored in the same size as the corresponding image. We treat the liver tumor regions as part of the liver in our experiments.

\textbf{Kidney dataset} contains 210 images, whose sizes range between [29$\sim$1059]$\times$[512]$\times$\{512,796\} (in $D$$\times$$H$$\times$$W$ pixels). The in-plane spacing of CT slices varies from 0.44$mm$ to 1.04$mm$ and the slice thickness varies from 0.50$mm$ to 5.00$mm$. Each CT image has a pixel-wise annotation of kidney and tumor segmentation stored in the same size as the corresponding image. We treat the kidney tumor regions as part of the kidney in our experiments.

\textbf{Pancreas dataset} contains 281 images, whose sizes range between [37$\sim$751]$\times$[512]$\times$[512] (in $D$$\times$$H$$\times$$W$ pixels). The in-plane spacing of these CT slices varies from 0.61$mm$ to 0.98$mm$ and the slice thickness varies from 0.70$mm$ to 7.50$mm$. Each CT image has a pixel-wise annotation of pancreas and tumor segmentation stored in the same size as the corresponding image. We treat the pancreas tumor regions as part of the pancreas in our experiments.

\textbf{Spleen dataset} contains 41 images, whose sizes range between [31$\sim$168]$\times$[512]$\times$[512] (in $D$$\times$$H$$\times$$W$ pixels). The in-plane spacing of these CT slices varies from 0.61$mm$ to 0.98$mm$ and the slice thickness varies from 1.50$mm$ to 8.00$mm$. Each CT image has a pixel-wise annotation of spleen segmentation stored in the same size as the corresponding image.

\textbf{AMOS dataset} contains 200 images, whose sizes range between [64$\sim$512]$\times$[60$\sim$768]$\times$[192$\sim$768] (in $D$$\times$$H$$\times$$W$ pixels). The in-plane spacing of these CT slices varies from 0.45$mm$ to 3.00$mm$ and the slice thickness varies from 0.82$mm$ to 5.00$mm$. Each CT image has a pixel-wise annotation of liver, kidney, pancreas, spleen, and gallbladder segmentation stored in the same size as the corresponding image.

\textbf{BTCV dataset} contains 30 images, whose sizes range between [85$\sim$198]$\times$[512]$\times$[512] (in $D$$\times$$H$$\times$$W$ pixels). The in-plane spacing of these CT slices varies from 0.59$mm$ to 0.98$mm$ and the slice thickness varies from 2.50$mm$ to 5.00$mm$. Each CT image has a pixel-wise annotation of  liver, kidney, pancreas, spleen, and gallbladder segmentation stored in the same size as the corresponding image.

For the liver, kidney, pancreas, spleen, and AMOS datasets, we randomly split each of them into training/validation/testing sets with a fixed ratio of 60\%:10\%:30\%, respectively. The experimental results on the five testing sets are used for \emph{in-federation} evaluation, which indicates the model performance when the testing data follows the same distribution as the training and validation data. For the BTCV dataset, we reserve it as an \emph{out-of-federation} testing set, which is completely unseen to the model during training and validation. The performance on the BTCV dataset gives a good indication of the model's generalization ability.  

\subsection{Metrics}
We calculated the mean and standard deviation (SD) of Dice similarity coefficient (DSC) and average symmetric surface distance (ASD) for each organ to quantitatively evaluate the segmentation results yielded by different methods. For each labeled organ in a certain client dataset, we first calculate the mean value $Q_c^k$ over all cases to get the result for organ $c$ in this client $k$. Then, the mean value of the $C_k$ labeled organs $Q^k=1/C_k \sum_{c=1}^{C_k}Q_c^k$ is calculated as the performance index of client $k$. Finally, the mean value of the $K$ clients $Q=1/K \sum_{k=1}^K Q^k$ is considered as the global index indicating the model accuracy. Instead of performing an overall case-wise averaging, this strategy was to avoid the bias from unbalanced sample numbers among the datasets, ensuring the global DSC and ASD calculated from each dataset with different sizes play an equal role. We also conducted paired $t$-tests on the above metrics to check the statistical significance between different groups of results.

\begin{table*}[t]
	\caption{Quantitative performance evaluation of different methods when tested on the in-federation data. The best results are marked in bold. The underlined results indicate a statistically significant difference from our result ({$p$}$<$0.05).}
	\centering
	\begin{threeparttable}[b]
		\begin{tabular}{p{60pt}<{\centering}|p{32pt}<{\centering}|p{32pt}<{\centering}|p{36pt}<{\centering}|p{32pt}<{\centering}|p{32pt}<{\centering}p{32pt}<{\centering}p{32pt}<{\centering}p{32pt}<{\centering}p{36pt}<{\centering}|p{24pt}<{\centering}}
			\toprule[2pt]
			\multirow{4}*{\makecell[c]{Models\\(\# of param.)}}
			~&\multicolumn{10}{c}{DSC [Mean\scriptsize{(SD)} \%]}\\
			\cmidrule[0.5pt]{2-11}
			~&Client~\#1&Client~\#2&Client~\#3&Client~\#4&\multicolumn{5}{c|}{Client~\#5}&\multirow{3}*{\makecell[c]{Global}}\\
			\cmidrule[0.5pt]{2-10}
            ~&Liver&Kidney&Pancreas&Spleen&Liver&Kidney&Pancreas&Spleen&Gallbladder\\
			\midrule[1pt]
            \multicolumn{1}{l|}{Localized}&&&&&&&&&\\
            \multicolumn{1}{l|}{- Client \#1}&93.46\scriptsize{(3.20)}&-&-&-&-&-&-&-&-\\
            \multicolumn{1}{l|}{- Client \#2}&-&\underline{91.17}\scriptsize{(7.60)}&-&-&-&-&-&-&-\\
            \multicolumn{1}{l|}{- Client \#3}&-&-&\underline{76.82}\scriptsize{(13.01)}&-&-&-&-&-&-&88.50\\
            \multicolumn{1}{l|}{- Client \#4}&-&-&-&\underline{91.98}\scriptsize{(3.75)}&-&-&-&-&-\\
            \multicolumn{1}{l|}{- Client \#5}&-&-&-&-&\underline{95.93}\scriptsize{(2.00)}&\underline{94.80}\scriptsize{(2.81)}&\underline{78.91}\scriptsize{(11.06)}&\underline{95.47}\scriptsize{(2.78)}&\underline{80.29}\scriptsize{(19.40)}\\
            \midrule[1pt]
            \multicolumn{1}{l|}{Centralized}&&&&&&&&&\\
            \multicolumn{1}{l|}{- U-Net~(97M)}&\underline{95.22}\scriptsize{(2.68)}&95.52\scriptsize{(3.30)}&80.50\scriptsize{(11.82)}&95.52\scriptsize{(1.82)}&96.69\scriptsize{(1.28)}&93.27\scriptsize{(4.48)}&82.25\scriptsize{(9.37)}&96.17\scriptsize{(2.95)}&84.76\scriptsize{(17.77)}&91.48\\
            \multicolumn{1}{l|}{- MENU-Net~(78M)}&\underline{\textbf{95.27}}\scriptsize{(2.81)}&95.52\scriptsize{(6.28)}&\underline{\textbf{81.48}}\scriptsize{(9.09)}&\textbf{96.32}\scriptsize{(1.58)}&\underline{\textbf{96.99}}\scriptsize{(1.00)}&\underline{\textbf{95.00}}\scriptsize{(2.33)}&\underline{\textbf{83.88}}\scriptsize{(8.67)}&\textbf{96.56}\scriptsize{(2.35)}&85.20\scriptsize{(17.98)}&\textbf{92.02}\\
            \midrule[1pt]
            \multicolumn{1}{l|}{Federated}&&&&&&&&&\\
            \multicolumn{1}{l|}{- U-Net~(97M)}&94.48\scriptsize{(2.28)}&\underline{94.54}\scriptsize{(6.18)}&79.38\scriptsize{(12.69)}&93.75\scriptsize{(4.71)}&96.76\scriptsize{(1.55)}&\underline{92.40}\scriptsize{(4.56)}&80.79\scriptsize{(11.89)}&\underline{95.72}\scriptsize{(2.97)}&\underline{83.38}\scriptsize{(15.62)}&90.39\\
            \multicolumn{1}{l|}{- MENU-Net~(78M)}&94.07\scriptsize{(3.09)}&\textbf{95.94}\scriptsize{(3.75)}&80.05\scriptsize{(12.02)}&94.65\scriptsize{(2.98)}&96.70\scriptsize{(1.15)}&93.74\scriptsize{(2.23)}&82.14\scriptsize{(9.45)}&96.34\scriptsize{(2.09)}&\textbf{86.08}\scriptsize{(14.48)}&91.14\\
			\bottomrule[1pt]
		\end{tabular}
	\end{threeparttable}
    \begin{threeparttable}[b]
		\begin{tabular}{p{60pt}<{\centering}|p{32pt}<{\centering}|p{32pt}<{\centering}|p{36pt}<{\centering}|p{32pt}<{\centering}|p{32pt}<{\centering}p{32pt}<{\centering}p{32pt}<{\centering}p{32pt}<{\centering}p{36pt}<{\centering}|p{24pt}<{\centering}}
			\toprule[1pt]
			\multirow{4}*{\makecell[c]{Models\\(\# of param.)}}
			~&\multicolumn{10}{c}{ASD [Mean\scriptsize{(SD)} mm]}\\
			\cmidrule[0.5pt]{2-11}
			~&Client~\#1&Client~\#2&Client~\#3&Client~\#4&\multicolumn{5}{c|}{Client~\#5}&\multirow{3}*{\makecell[c]{Global}}\\
			\cmidrule[0.5pt]{2-10}
            ~&Liver&Kidney&Pancreas&Spleen&Liver&Kidney&Pancreas&Spleen&Gallbladder\\
			\midrule[1pt]
            \multicolumn{1}{l|}{Localized}&&&&&&&&&\\
            \multicolumn{1}{l|}{- Client \#1}&\underline{4.00}\scriptsize{(2.88)}&-&-&-&-&-&-&-&-\\
            \multicolumn{1}{l|}{- Client \#2}&-&\underline{2.40}\scriptsize{(4.26)}&-&-&-&-&-&-&-\\
            \multicolumn{1}{l|}{- Client \#3}&-&-&\underline{3.01}\scriptsize{(3.82)}&-&-&-&-&-&-&2.34\\
            \multicolumn{1}{l|}{- Client \#4}&-&-&-&\underline{1.11}\scriptsize{(0.62)}&-&-&-&-&-\\
            \multicolumn{1}{l|}{- Client \#5}&-&-&-&-&0.95\scriptsize{(1.08)}&0.87\scriptsize{(3.31)}&1.48\scriptsize{(1.03)}&\underline{0.67}\scriptsize{(1.01)}&\underline{1.99}\scriptsize{(3.33)}&~\\
            \midrule[1pt]
            \multicolumn{1}{l|}{Centralized}&&&&&&&&&\\
            \multicolumn{1}{l|}{- U-Net~(97M)}&\underline{1.95}\scriptsize{(1.63)}&\underline{3.26}\scriptsize{(6.31)}&2.04\scriptsize{(2.80)}&1.17\scriptsize{(1.62)}&1.02\scriptsize{(1.82)}&\underline{4.87}\scriptsize{(9.15)}&\underline{1.30}\scriptsize{(1.26)}&0.64\scriptsize{(1.93)}&1.52\scriptsize{(3.26)}&2.06\\
            \multicolumn{1}{l|}{- MENU-Net~(78M)}&\underline{\textbf{1.86}}\scriptsize{(2.07)}&0.88\scriptsize{(1.98)}&\textbf{1.85}\scriptsize{(1.49)}&\textbf{0.30}\scriptsize{(0.13)}&\textbf{0.61}\scriptsize{(0.47)}&\textbf{0.72}\scriptsize{(1.40)}&\underline{\textbf{1.19}}\scriptsize{(1.28)}&\textbf{0.36}\scriptsize{(0.62)}&\textbf{1.24}\scriptsize{(2.22)}&\textbf{1.14}\\
            \midrule[1pt]
            \multicolumn{1}{l|}{Federated}&&&&&&&&&\\
            \multicolumn{1}{l|}{- U-Net~(97M)}&3.75\scriptsize{(4.08)}&\underline{3.11}\scriptsize{(6.11)}&2.36\scriptsize{(4.01)}&1.54\scriptsize{(2.72)}&\underline{1.40}\scriptsize{(1.86)}&\underline{2.82}\scriptsize{(6.02)}&1.93\scriptsize{(3.87)}&0.45\scriptsize{(0.60)}&1.56\scriptsize{(2.89)}&2.48\\
            \multicolumn{1}{l|}{- MENU-Net~(78M)}&2.56\scriptsize{(2.47)}&\textbf{0.79}\scriptsize{(1.03)}&2.10\scriptsize{(2.45)}&0.54\scriptsize{(0.42)}&0.94\scriptsize{(1.59)}&0.80\scriptsize{(0.89)}&1.53\scriptsize{(1.54)}&0.41\scriptsize{(0.52)}&1.34\scriptsize{(2.70)}&1.40\\
			\bottomrule[2pt]
		\end{tabular}
	\end{threeparttable}
	\label{tab:2}
\end{table*}

\subsection{Comparison with benchmarks}
\label{sec:comparison}

\subsubsection{Benchmarks}
We compared our method with three benchmarks to demonstrate its effectiveness. The benchmarks include:

\textbf{Localized learning mode}: Four single-organ segmentation networks and a five-organ segmentation network were individually trained using the liver, kidney, pancreas, spleen, and AMOS datasets, respectively. This benchmark simulated the scenario where the clinical sites cannot share their data with each other and no techniques are adopted to deal with the data privacy issue during the model training.

\textbf{Centralized learning mode}: A five-organ segmentation network was trained using the combination of the liver, kidney, pancreas, spleen, and AMOS datasets. This benchmark simulated an ideal scenario where the clinical sites can freely share their data without any concern on the data privacy during the model training.

\textbf{Federated learning mode}: A five-organ segmentation network was trained using the FedAvg~\cite{Mcmahan2017FedAvg} algorithm with the liver, kidney, pancreas, spleen, and AMOS datasets distributed on five client nodes. This benchmark simulated a practical scenario where the clinical sites cannot share their data with each other and a naive FL solution is adopted to deal with the data privacy issue during the model training.

We adopted U-Net~\cite{Ronneberger2015UNet} as the segmentation network for the above models. To ensure the performance gain comes from the design choice rather than the additional network parameters, we extended the base channel number of the U-Net from 32 to 96 to make it have a slightly larger number of parameters than our method. In the leftmost column of Table II and Table III, we specified the trainable parameters in each model. The training configuration of all the methods followed the recommended settings by nnU-Net~\cite{Isensee2021nnUNet}. For the U-Net in federated learning mode, we employed the marginal loss and exclusion loss~\cite{Shi2021MarginalLoss} as the training objective to adapt it to the partially labeled client data. The core idea of the marginal/exclusion loss is to combine the unlabeled organs with the background category and try to maximize/minimize the overlap with the merged background/foreground ground truth, respectively, which can be formulated as follows:
\begin{align} 
    L_{margin} &= L_{ce}(p', 1-\hat{y}') + L_{dice}(p', 1-\hat{y}')\\
    L_{exclusion} &= -L_{ce}(p', \hat{y}') - L_{dice}(p', \hat{y}')
\end{align}
where $p'$ and $\hat{y}'$ denote the predicted segmentation of the merged background category and ground-truth segmentation of the merged foreground category, respectively. We utilized the sum of the Cross-entropy loss and Dice loss~\cite{Milletari2016VNet} to quantify the overlapping regions, which is consistent with the original nnU-Net~\cite{Isensee2021nnUNet} training settings. For the sake of fairness, the same objective was also used as the supervised loss $L_{sup}$ in our method.

\begin{table*}[t]
	\caption{Quantitative performance evaluation of different methods when tested on the out-of-federation data. The best results are marked in bold. The underlined results indicate a statistically significant difference from our result ({$p$}$<$0.05).}
	\centering
    \scalebox{0.85}{
	\begin{threeparttable}[b]
		\begin{tabular}{p{60pt}<{\centering}|p{34pt}<{\centering}p{34pt}<{\centering}p{34pt}<{\centering}p{34pt}<{\centering}p{34pt}<{\centering}|p{24pt}<{\centering}|p{32pt}<{\centering}p{32pt}<{\centering}p{32pt}<{\centering}p{32pt}<{\centering}p{36pt}<{\centering}|p{22pt}<{\centering}}
			\toprule[2pt]
			\multirow{3}*{\makecell[c]{Models\\(\# of param.)}}
			~&\multicolumn{6}{c|}{DSC [Mean\scriptsize{(SD)} \%]}&\multicolumn{6}{c}{ASD [Mean\scriptsize{(SD)} mm]}\\
			\cmidrule[0.5pt]{2-13}
			~&Liver&Kidney&Pancreas&Spleen&Gallbladder&\multicolumn{1}{c|}{Global}&Liver&Kidney&Pancreas&Spleen&Gallbladder&Global\\
			\midrule[1pt]
            \multicolumn{1}{l|}{Localized}&&&&&&&&&&&&\\
            \multicolumn{1}{l|}{- Client \#1}&\underline{92.63}\scriptsize{(3.46)}&-&-&-&-&~&\underline{3.70}\scriptsize{(4.93)}&-&-&-&-&~\\
            \multicolumn{1}{l|}{- Client \#2}&-&\underline{85.07}\scriptsize{(16.53)}&-&-&-&~&-&5.28\scriptsize{(17.98)}&-&-&-&~\\
            \multicolumn{1}{l|}{- Client \#3}&-&-&\underline{73.78}\scriptsize{(11.47)}&-&-&76.13&-&-&6.02\scriptsize{(7.19)}&-&-&4.57\\
            \multicolumn{1}{l|}{- Client \#4}&-&-&-&\underline{85.01}\scriptsize{(15.57)}&-&~&-&-&-&3.64\scriptsize{(9.40)}&-&~\\
            \multicolumn{1}{l|}{- Client \#5}&\underline{92.78}\scriptsize{(4.97)}&\textbf{89.14}\scriptsize{(16.38)}&\underline{57.13}\scriptsize{(26.67)}&88.18\scriptsize{(15.80)}&\underline{53.45}\scriptsize{(33.93)}&~&\underline{3.15}\scriptsize{(5.41)}&\textbf{4.19}\scriptsize{(11.60)}&6.73\scriptsize{(7.94)}&\underline{3.49}\scriptsize{(5.60)}&\underline{5.29}\scriptsize{(9.17)}&~\\
            \midrule[1pt]
            \multicolumn{1}{l|}{Centralized}&&&&&&&&&&&&\\
            \multicolumn{1}{l|}{- U-Net~(97M)}&95.35\scriptsize{(4.07)}&\underline{85.50}\scriptsize{(16.97)}&77.44\scriptsize{(11.05)}&91.46\scriptsize{(10.59)}&68.02\scriptsize{(32.17)}&83.55&2.06\scriptsize{(4.68)}&\underline{9.09}\scriptsize{(15.33)}&2.91\scriptsize{(3.97)}&2.49\scriptsize{(5.23)}&2.08\scriptsize{(3.43)}&3.73\\
            \multicolumn{1}{l|}{- MENU-Net~(78M)}&\underline{\textbf{95.84}}\scriptsize{(3.35)}&88.70\scriptsize{(15.30)}&\textbf{78.65}\scriptsize{(9.98)}&\textbf{92.57}\scriptsize{(8.89)}&67.86\scriptsize{(31.26)}&\textbf{84.72}&\underline{\textbf{1.85}}\scriptsize{(4.57)}&4.26\scriptsize{(14.35)}&\textbf{2.48}\scriptsize{(4.15)}&1.58\scriptsize{(3.38)}&3.79\scriptsize{(13.25)}&2.79\\
            \midrule[1pt]
            \multicolumn{1}{l|}{Federated}&&&&&&&&&&&&\\
            \multicolumn{1}{l|}{- U-Net~(97M)}&95.00\scriptsize{(5.41)}&88.03\scriptsize{(15.76)}&76.53\scriptsize{(10.52)}&92.37\scriptsize{(12.44)}&64.35\scriptsize{(30.94)}&83.26&2.76\scriptsize{(6.66)}&5.97\scriptsize{(15.98)}&2.59\scriptsize{(2.71)}&\textbf{1.13}\scriptsize{(2.39)}&\textbf{1.70}\scriptsize{(1.87)}&2.83\\
            \multicolumn{1}{l|}{- MENU-Net~(78M)}&95.21\scriptsize{(4.11)}&88.12\scriptsize{(16.88)}&77.92\scriptsize{(10.91)}&90.97\scriptsize{(13.29)}&\textbf{69.45}\scriptsize{(31.61)}&84.33&2.39\scriptsize{(5.74)}&4.56\scriptsize{(14.78)}&3.44\scriptsize{(9.85)}&1.38\scriptsize{(2.64)}&1.79\scriptsize{(3.28)}&\textbf{2.71}\\
			\bottomrule[2pt]
		\end{tabular}
	\end{threeparttable}}
	\label{tab:3}
\end{table*}

\subsubsection{In-federation performance}

Table~\ref{tab:2} shows the quantitative results of the in-federation evaluation. It can be observed that, among the three learning modes, the centralized learning models generally exhibited the upper-bound performance. The FL models significantly outperformed the localized learning models, which justified the feasibility of using different partially labeled datasets to collaboratively train a multi-organ segmentation model through FL. When compared with the baseline U-Net model, our method achieved higher global accuracy in both centralized learning mode and FL mode, which demonstrated the superiority of our designs. Although not all the improvement margins are statistically significant, the standard deviation of our results is much lower than those of the U-Net baseline model, which indicated our method had a more stable performance on the multi-center data. We also observed that, while the region-based DSC metric showed relatively small differences, our model performed much better than the U-Net in terms of the distance-based ASD metric. This phenomenon suggested that there are fewer false positives in our results. It is worth noting that the trainable parameter number of our method is $\sim$78 million, which is 24\% fewer than that of the U-Net ($\sim$97 million parameters). This result suggests that the performance gain of our method was coming from the design choice rather than the additional parameters. More importantly, the smaller model size can also benefit the FL due to the lower communication and computation costs.

\subsubsection{Out-of-federation performance}
Table~\ref{tab:3} presents the quantitative results of the out-of-federation evaluation. Similar to the in-federation scenario, the centralized learning models showed superior accuracy over the FL and localized learning models. However, there was a global performance degradation from the in-federation results to the out-of-federation results due to the data distribution shift from the training domain (the liver, kidney, pancreas, spleen, and AMOS datasets) to an unseen testing domain (the BTCV dataset). Furthermore, we observed that the performance gap between the centralized learning models and the FL models decreased for the out-of-federation results, as compared to the in-federation results. These results suggest that out-of-federation segmentation is a more challenging task than in-federation segmentation, and our method can effectively handle it with better performance over the baseline U-Net model.

\begin{figure*}
	\centering
	\includegraphics[width=0.95\textwidth]{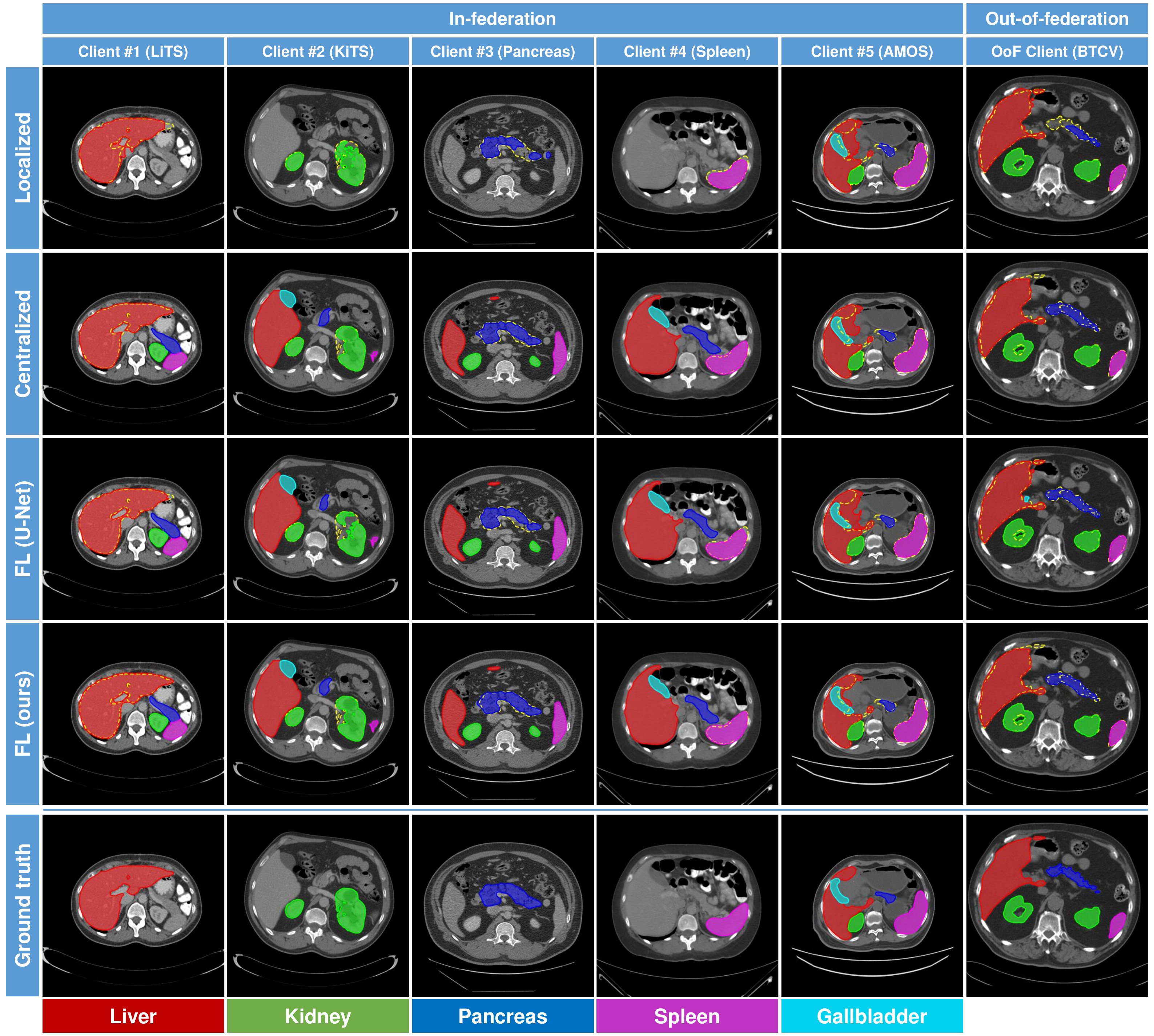}
	\caption{2D Visualization of segmentation results yielded by the competing methods. Each column shows a case from one client. For better comparison, the ground-truth contours are also superimposed on the segmentation results as yellow dashed lines.}
	\label{fig:2}
\end{figure*}

\subsubsection{Result visualization}
\begin{figure*}
	\centering
	\includegraphics[width=0.95\textwidth]{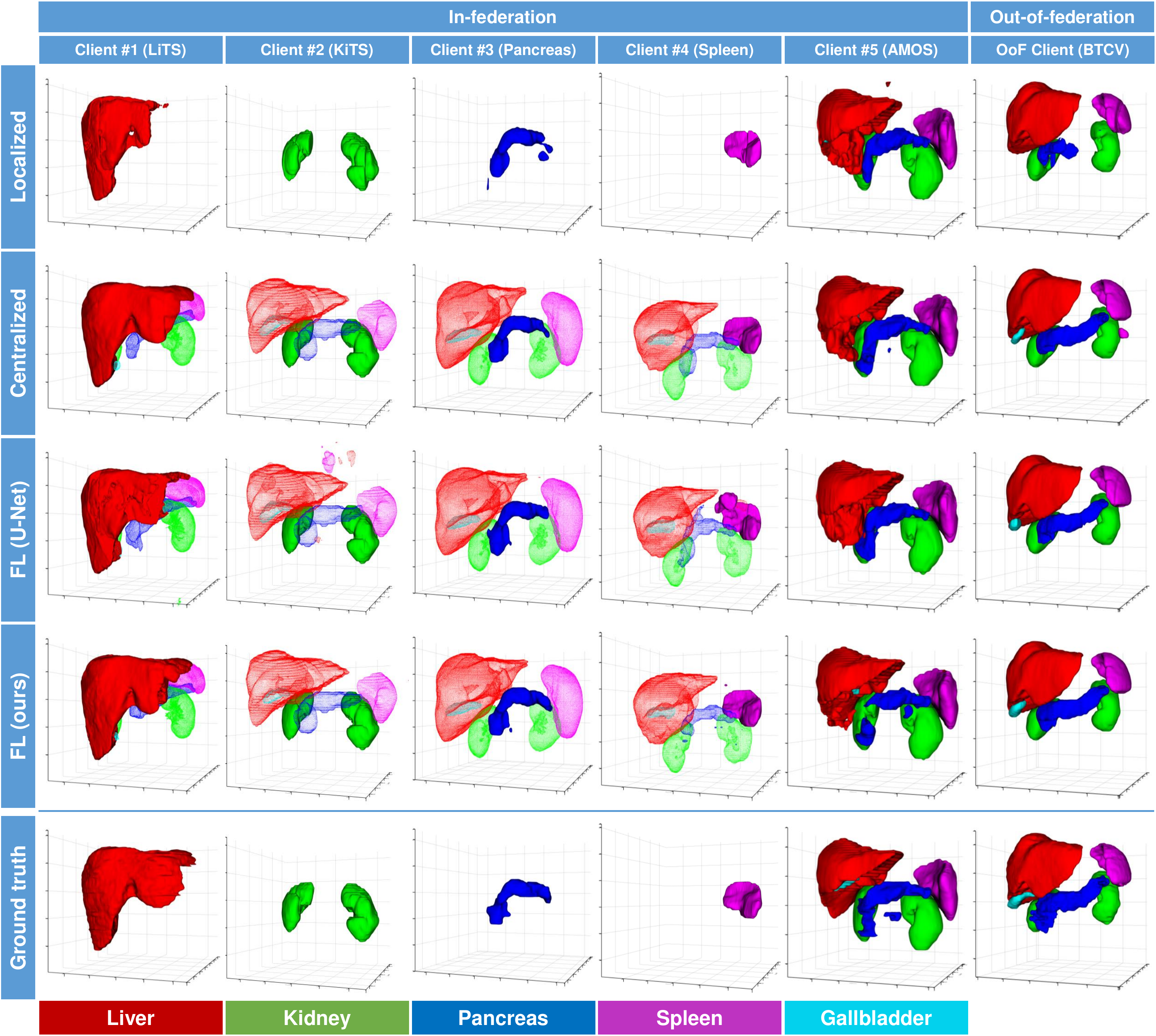}
	\caption{3D Visualization of segmentation results by the competing methods. Each column shows a case from one client. The segmentation of the unlabeled organs is represented by transparent meshes.}
	\label{fig:3}
\end{figure*}

Figs.~\ref{fig:2} and \ref{fig:3} visualize the segmentation results yielded by different methods in 2D (axial) and 3D views, respectively. Each column shows a case from one client dataset. In the 2D views, the ground-truth contours are superimposed on the segmentation results as yellow dashed lines for better comparison. In the 3D views, the segmentation results of the unlabeled organs are represented by transparent meshes. It can be observed that our proposed method produced fewer false positive results in either in-federation or out-of-federation results.

\begin{table}[t]
	\caption{Experimental results of ablation studies on the proposed method. The best results are marked in bold.}
	\centering
	\begin{threeparttable}[b]
		\begin{tabular}{p{60pt}|p{24pt}<{\centering}p{32pt}<{\centering}|p{24pt}<{\centering}p{32pt}<{\centering}}
			\toprule[2pt]
			\multirow{4}*{\makecell[c]{Ablation models}}
			~&\multicolumn{2}{c|}{In-federation}&\multicolumn{2}{c}{Out-of-federation}\\
			\cmidrule[0.5pt]{2-5}
			~&\makecell[c]{DSC [\%]}&\makecell[c]{ASD [mm]}&\makecell[c]{DSC [\%]}&\makecell[c]{ASD [mm]}\\
			\midrule[1pt]
            \multicolumn{1}{l|}{baseline}&90.39&2.48&83.26&2.83\\
            \multicolumn{1}{l|}{+MENU-Net}&90.95&1.76&83.86&3.39\\
            \multicolumn{1}{l|}{+NEMU-Net+ALD}&90.78&1.62&83.36&3.65\\
            \multicolumn{1}{l|}{+NEMU-Net+AGD}&\textbf{91.14}&\textbf{1.40}&\textbf{84.33}&\textbf{2.71}\\
			\bottomrule[1pt]
		\end{tabular}
	\end{threeparttable}
	\label{tab:4}
\end{table}

\begin{figure}
	\centering
	\includegraphics[width=\columnwidth]{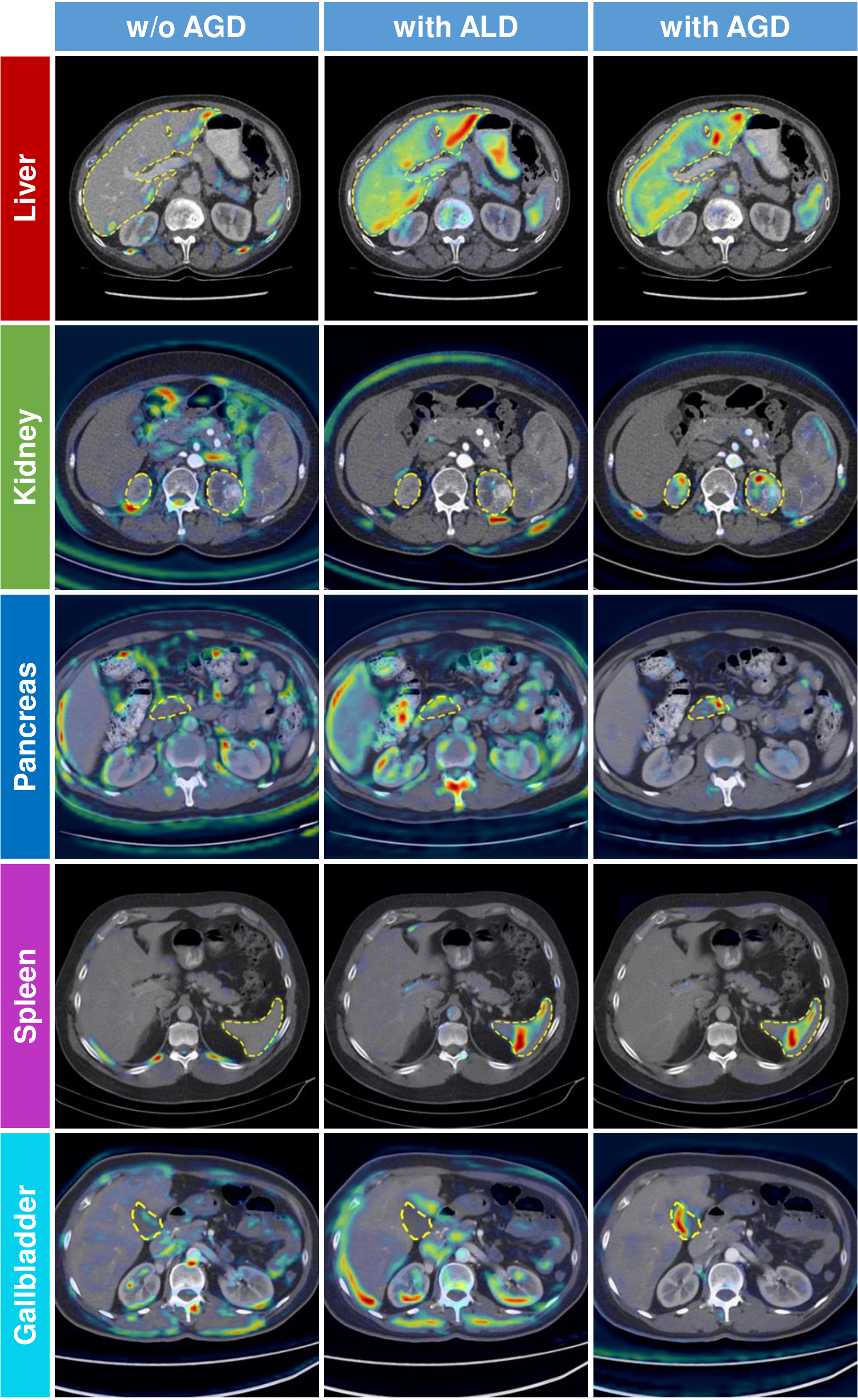}
	\caption{Gradient-weighted class activation maps (Grad-CAMs)~\cite{Selvaraju2017GradCAM} generated by the MENU-Net trained without AGD (left), with ALD (middle), and with AGD (right). Yellow dashed lines indicate ground-truth contours. We use the activation (output) of the third convolutional block in the sub-encoders to calculate the Grad-CAMs.}
	\label{fig:4}
\end{figure}

\subsection{Ablation study}
\label{sec:experiments:ablation}
In this section, we conducted ablation studies on the proposed method to evaluate the effectiveness of the two key designs, including 1) the MENU-Net for organ-specific feature extraction and 2) the AGD for organ-specific feature enhancement. A U-Net~\cite{Ronneberger2015UNet} trained using the FedAvg algorithm~\cite{Mcmahan2017FedAvg} was used as the baseline method, on which the MENU-Net and AGD components are sequentially imposed. In addition, we also tried to replace the AGD with a non-generic version, \emph{i.e.}, the auxiliary decoders are locally trained on the client nodes without parameter fusion (see Eq.~\ref{eq:1}) through the server node. We denote it as \emph{auxiliary localized decoder} (ALD) in the following contents. This ablation model is used to justify the necessity of parameter sharing in learning organ-specific features.

The quantitative results of the ablation studies are listed in Table~\ref{tab:4}. The observations are three-fold. 1) Compared with the baseline, the design of the MENU-Net (Table~\ref{tab:4}, ``\textbf{+MENU-Net}'') can effectively improve the global segmentation indices except for the ASD metric  in out-of-federation evaluation. This result indicated a positive effect of the task decomposition design (\emph{i.e.}, the multi-encoder network architecture) in learning domain-invariant features.
2) Based on the MENU-Net, the AGD (Table~\ref{tab:4}, ``\textbf{+MENU-Net+AGD}'') can further boost the segmentation accuracy for both in-federation and out-of-federation evaluation. This result suggested that the auxiliary supervisions on the intermediate layers can facilitate the learning of more discriminative features, and thus, benefit the subsequent shared decoder for accurate segmentation. 3) By replacing the AGD with the ALD, the accuracy of our method (Table~\ref{tab:4}, ``\textbf{+MENU-Net+ALD}'') dropped in both the in-federation setting and the out-of-federation setting. This result demonstrated the importance of parameter sharing in our AGD since the auxiliary decoder shared across the clients enforced the preceding sub-encoders to extract discriminative features for different organs.

\begin{table}[t]
	\caption{Quantitative performance evaluation of the proposed method trained with different communication frequencies $T$$\times$$E$. $T$ is the number of communication rounds and $E$ is the number of local training epochs. The best results are marked in bold.}
	\centering
    \begin{threeparttable}[b]
		\begin{tabular}{p{60pt}|p{24pt}<{\centering}p{32pt}<{\centering}|p{24pt}<{\centering}p{32pt}<{\centering}}
			\toprule[2pt]
			\multirow{5}*{\makecell[c]{Communication\\frequency\\$T$ $\times$ $E$}}
			~&\multicolumn{2}{c|}{In-federation}&\multicolumn{2}{c}{Out-of-federation}\\
			\cmidrule[0.5pt]{2-5}
			~&\makecell[c]{DSC [\%]}&\makecell[c]{ASD [mm]}&\makecell[c]{DSC [\%]}&\makecell[c]{ASD [mm]}\\
			\midrule[1pt]
            \multicolumn{1}{c|}{25$\times$16}&90.63&1.78&83.10&3.47\\
            \multicolumn{1}{c|}{50$\times$8}&90.64&1.92&83.45&2.95\\
            \multicolumn{1}{c|}{100$\times$4}&90.72&1.89&83.58&2.98\\
            \multicolumn{1}{c|}{200$\times$2}&90.95&1.57&83.81&3.27\\
            \multicolumn{1}{c|}{400$\times$1}&\textbf{91.14}&\textbf{1.40}&\textbf{84.33}&\textbf{2.71}\\
			\bottomrule[1pt]
		\end{tabular}
	\end{threeparttable}
	\label{tab:5}
\end{table}

Fig.~\ref{fig:4} visualized the gradient-weighted class activation maps (Grad-CAMs)~\cite{Selvaraju2017GradCAM} produced by our MENU-Net (extracted from the third convolutional block in the sub-encoders) when it is trained without AGD, with ALD, and with AGD, respectively. The model trained with AGD generated activation maps with more accurate shapes and fewer false positive regions than the other two models, which indicated that the AGD can effectively enhance the organ-specific features extracted by the sub-encoders with better explainability~\cite{Yang2022XAI}. 

\begin{figure}
	\centering
	\includegraphics[width=\columnwidth]{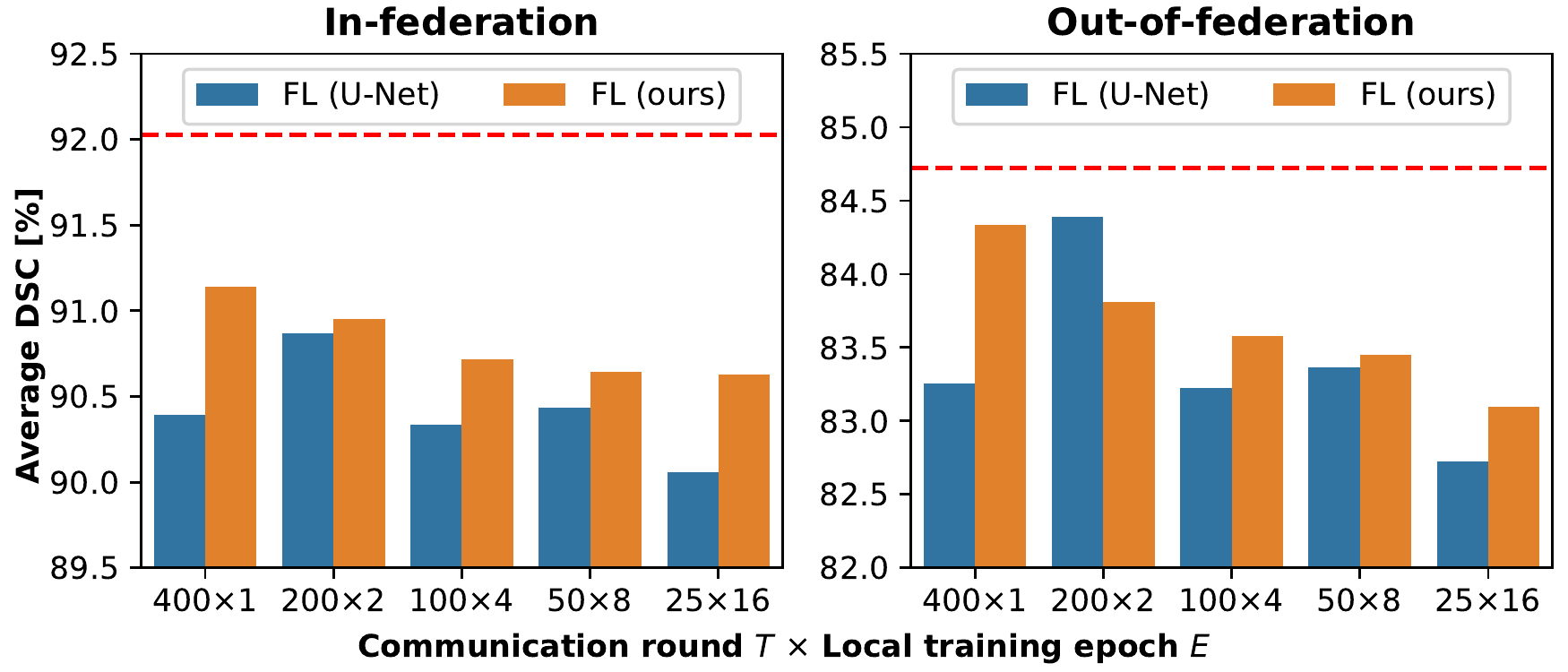}
	\caption{Average DSC of FL models trained with different communication frequencies. The red dashed line indicates the average DSC achieved by centralized learning model.}
	\label{fig:5}
\end{figure}

\subsection{Effects of communication frequency}
Communication frequency is a key factor affecting the performance of the FL methods in practice. In the proposed method, the communication frequency is jointly controlled by the number of communication rounds $T$ and the number of local training epochs $E$. We successively trained our method with different combinations of $T$$\times$$E$ to investigate the effects of communication frequency. For the sake of fairness, the product of these two parameters are fixed to 400, which means all models are optimized with the same number of training batches (iterations).

The experimental results are shown in Table~\ref{tab:5} and Fig.~\ref{fig:5}. It can be seen that higher communication frequency generally brought better performance to the trained model, which is in line with observations in other FL-based methods~\cite{Mcmahan2017FedAvg,Xia2021AutoFedAvg,Li2021FedBN,Yang2021FedSemiCovid}, and our method consistently outperformed the baseline FL U-Net model. Considering our method is designed for the cross-silo FL~\cite{Kairouz2021APinFL} scenario, where the clients (clinical sites) have stable internet connections with sufficient bandwidth, we finally choose $T$=400 and $E$=1 in our method to achieve higher accuracy.

\begin{table}[t]
	\caption{Quantitative performance evaluation of the proposed method when combined with different federated learning strategies. The best results are marked in bold.}
	\centering
	\begin{threeparttable}[b]
		\begin{tabular}{p{60pt}|p{24pt}<{\centering}p{24pt}<{\centering}|p{24pt}<{\centering}p{24pt}<{\centering}}
			\toprule[2pt]
			\multirow{4}*{\makecell[c]{Models}}
			~&\multicolumn{2}{c|}{In-federation}&\multicolumn{2}{c}{Out-of-federation}\\
			\cmidrule[0.5pt]{2-5}
			~&DSC [\%]&ASD [mm]&DSC [\%]&ASD [mm]\\
			\midrule[1pt]
            \multicolumn{1}{l|}{FedAvg~\cite{Mcmahan2017FedAvg}}&91.14&\textbf{1.40}&84.33&2.71\\
            \multicolumn{1}{l|}{FedAvgM~\cite{Hsu2019FedAvgM}}&90.28&1.98&82.32&3.29\\
            \multicolumn{1}{l|}{FedProx~\cite{Li2020FedProx}}&90.95&1.62&83.73&2.97\\
            \multicolumn{1}{l|}{FedDyn~\cite{Acar2021FedDyn}}&\textbf{91.33}&1.56&\textbf{84.60}&3.23\\
            FedSM~\cite{Xu2022FedSM}&91.15&1.41&84.38&\textbf{2.70}\\
            FedNorm~\cite{Bernecker2022FedNorm}&90.92&1.75&84.21&3.12\\
            \multicolumn{1}{l|}{Fed-DoDNet~\cite{Zhang2021PartialLabel}}&91.09&1.75&83.82&2.74\\
			\bottomrule[1pt]
		\end{tabular}
	\end{threeparttable}
	\label{tab:6}
\end{table}

\subsection{Comparison with different FL strategies}
Another important factor relating to the FL systems’ performance is the choice of the learning strategy, which determines how the locally trained models are optimized and aggregated to update the server model. In the prior experiments, to focus on our own designs, we consistently used the most basic FL strategy, \emph{i.e.}, the FedAvg algorithm~\cite{Mcmahan2017FedAvg}, for this choice. Our method is also flexible to be combined with other FL strategies. To demonstrate this property, we tried to deploy our method with six different FL strategies, including FedAvg~\cite{Mcmahan2017FedAvg}, FedAvgM~\cite{Hsu2019FedAvgM}, FedProx~\cite{Li2020FedProx}, FedDyn~\cite{Acar2021FedDyn}, FedSM~\cite{Xu2022FedSM}, and FedNorm~\cite{Bernecker2022FedNorm}. The experimental results of these hybrid models are shown in Table~\ref{tab:6}. It can be observed that, by combining with more advanced FL strategies, the segmentation accuracy of our method can be further improved. The highest DSC was achieved when our method was combined with the FedDyn~\cite{Acar2021FedDyn} algorithm. This experiment demonstrates the flexibility of our proposed method. While the existing FL strategies mainly focus on resolving the problem of image heterogeneities, our method is designed to deal with label heterogeneities. Thus, the proposed method and FL strategies can complement each other. We also compared our method with another multi-organ segmentation method for partially labeled data, \emph{i.e.}, DoDNet~\cite{Zhang2021PartialLabel}, which was originally designed for centralized learning scenario and extended in this experiment by combining with the FedAvg~\cite{Mcmahan2017FedAvg} algorithm (denoted as ``\textbf{Fed-DoDNet}'' in Table~\ref{tab:6}). The segmentation accuracy of this competing model is slightly lower than our model, which demonstrated the superiority of our method designs.

\section{Discussion and conclusion}
\label{sec:conclusion}
In our experiments, the FL model was trained with five client datasets, each containing varying numbers of images that were labeled with different organs. This setup was designed to simulate real scenarios in clinical settings. Our experimental results revealed that our proposed method is flexible and capable of handling such complex scenarios. However, it is important to note that our approach has a minimum criterion for the number of labeled organs required. Specifically, each client dataset must be labeled with at least one of the target organs. For more complicated scenarios involving unlabeled images, it would require some additional semi-supervised or unsupervised learning techniques, such as the teacher-student consistency learning strategy~\cite{Tarvainen2017MeanTeacher,Xu2021Shadow}, to deal with. Fortunately, our Fed-MENU framework is highly adaptable and can seamlessly integrate these methods, making it possible to extend the framework to these situations.

\begin{table}[t]
	\caption{Quantitative performance evaluation of the proposed method when the lesion regions were treated as background/a part of the target organs. The best results are marked in bold.}
	\centering
	\begin{threeparttable}[b]
		\begin{tabular}{p{60pt}|p{24pt}<{\centering}p{24pt}<{\centering}|p{24pt}<{\centering}p{24pt}<{\centering}}
			\toprule[2pt]
			\multirow{4}*{\makecell[c]{Models}}
			~&\multicolumn{2}{c|}{In-federation}&\multicolumn{2}{c}{Out-of-federation}\\
			\cmidrule[0.5pt]{2-5}
			~&DSC [\%]&ASD [mm]&DSC [\%]&ASD [mm]\\
			\midrule[1pt]
            \multicolumn{1}{l|}{Lesion as background}&89.76&1.65&84.00&2.81\\
            \multicolumn{1}{l|}{Lesion as part of organ}&\textbf{91.14}&\textbf{1.40}&\textbf{84.33}&\textbf{2.71}\\
			\bottomrule[1pt]
		\end{tabular}
	\end{threeparttable}
	\label{tab:7}
\end{table}

Most of the CT images used in this study contained tumors, and we treated the lesion regions as a part of the surrounding organ in the experiments. There could be a potential influence of these abnormities on the performance of the proposed method. To investigate this impact, we compared the performance of our method when the lesion regions in the CT images were treated 1) as background and 2) as part of the foreground organs, respectively. The quantitative results yielded from these two settings are reported in Table~\ref{tab:7}. It can be seen that, when the lesion regions were treated as the background category, the resulting performance of the proposed method degraded. The reason can be attributed to the fact that, although the lesion regions exhibited different textures or intensities than the healthy tissues, they may not substantially change the contour shape of the target organs. Thus, it would be relatively easy to segment these regions as a part of the foreground. Otherwise, if the lesions were treated as the background, it may significantly change the structure of the target organs, which would eventually affect the segmentation accuracy. Since the proposed method is designed to deal with the inconsistently labeled data for FL, we can also consider the tumor labels as a new class inconsistently distributed in different client nodes and train our method to segment the tumors. This could lead us to a potential application of disease diagnosis. Naturally, to be adaptive to the task of diagnosis (classification), the output branch of the proposed MENU-Net may be replaced with a global pooling layer followed by fully-connected layers. As this study mainly focused on the application of multi-organ segmentation, this direction will be explored in our future research.

The experiments conducted in this study are done in a virtual environment, in which we focus on the training problem of inconsistently labeled data. In practical scenarios, there are some security issues that should be considered. For example, encryptions are required to secure the communications between server and client nodes. Unencrypted communication would raise a risk of information leakage and interference from malicious attacks~\cite{Kaissis2020DP}. In addition, challenges of data memorization~\cite{Song2017Memorization} as well as adversarial inference~\cite{Usynin2021AdvInfer} also raise privacy concerns to the FL methods. It has been demonstrated that an FL model can be maliciously manipulated to reconstruct the private training data if no protective measures are taken~\cite{Kaissis2021DP}. To address this issue, differential privacy (DP)~\cite{Dwork2014DP} techniques can be used to enhance the FL methods by introducing noise to their input data, computation results, or optimization process~\cite{Rajkumar2012DP}. Since the goal of this study is to develop an FL framework to enable collaborative learning on inconsistently labeled data, we did not draw extra efforts to the security issue. Fortunately, our method is compatible to the above-mentioned techniques and thus can be combined with them to achieve secure FL.

One potential limitation of the proposed method could be its scalability. In this study, we designed our method for a five-organ segmentation task. However, when more organs are desired in the segmentation task, more encoders are required in the MENU-Net. As a consequence, the concatenation of the output feature maps from the multi-encoders would increase model complexity and GPU memory consumption. One solution to this issue could be using one single shared encoder combined with conditional input to replace the multi-encoder design. This would result in a variant of our method sharing the DoDNet structure~\cite{Zhang2021PartialLabel}. In Table\ref{tab:6}, we reported the experimental results of this variant model (denoted as ``\textbf{Fed-DoDNet}''), which demonstrated the feasibility of this “shared-encoder+conditional input” design.

In this paper, we revealed and defined a new problem of FL with partially labeled data in the context of medical image segmentation, which is of great clinical significance and technical urgency to solve. Subsequently, a novel Fed-MENU method was proposed to tackle this challenging problem. Compared with the conventional FL framework that worked on the fully-labeled data, our Fed-MENU method had two key designs to adapt to the partially labeled client datasets, 1) the MENU-Net for organ-specific feature extraction and 2) the AGD for organ-specific feature enhancement. Extensive experiments were conducted using six public abdominal CT image datasets. The experimental results comprehensively demonstrated the feasibility and effectiveness of our method in solving the partial label problem in the context of FL.

\bibliographystyle{IEEEtran}
\bibliography{main}

\end{document}